\documentclass[cameraready]{Interspeech}

\title{QC-GAN: A Parameter-Efficient Quaternion Conformer GAN for High-Fidelity Speech Enhancement}

\author[affiliation={1}]{Shogo}{Yamauchi}
\author[affiliation={1}]{Hideaki}{Tamori}
\author[affiliation={1}]{Makoto}{Sakai}
\author[affiliation={1}]{Yosuke}{Yamano}
\author[affiliation={2}]{Tohru}{Nitta}

\address{
    $^1$ The Asahi Shimbun Company, Japan \\
    $^2$ Tokyo Woman's Christian University, Japan
}

\email{
    yamauchi-s1@asahi.com
}


\usepackage{comment}
\usepackage{amsmath}
\usepackage{hyperref}
\usepackage{cleveref}

\usepackage{microtype}
\begin{document}

\maketitle

\begin{abstract}
We propose a parameter-efficient speech enhancement framework, Quaternion Conformer GAN (QC-GAN), which combines a Quaternion Conformer generator with MetricGAN-based training. The Hamilton product encodes the magnitude and phase via structured weight sharing, reducing the number of layer parameters while preserving their interdependencies. A metric-learning discriminator was employed to maximize perceptual quality by optimizing the approximate perceptual evaluation scores. On the VoiceBank+DEMAND dataset, QC-GAN achieved a Perceptual Evaluation of Speech Quality (PESQ) score of 3.48 with only 0.89M parameters, delivering a performance comparable to state-of-the-art models at less than half their size. A 35K-parameter variant achieved a PESQ score of 3.23, surpassing conventional methods with significantly fewer parameters. Evaluation on the DNS-Challenge 3 dataset further confirmed generalization to real-world conditions. \footnote{https://github.com/asahi-research/QC-GAN}
\end{abstract}

\section{Introduction}
Speech enhancement (SE) is a critical task that aims to improve the perceptual quality of speech signals in real-world environments. With recent advances in deep learning, methods operating in the time-frequency (T--F) domain have become a dominant approach, resulting in significant performance improvements. In particular, Transformers with self-attention mechanisms~\cite{vaswani2017attention}, Conformers~\cite{gulati2020conformer} that combine Transformers with convolutional neural networks (CNNs), and state-space models~\cite{chao2024} have catalyzed a significant paradigm shift in this field. Conformer-based architectures, in particular, have achieved state-of-the-art (SoTA) performances through models such as CMGAN~\cite{Cao2022} and MP-SENet~\cite{Lu_2023}, establishing themselves as foundational frameworks for SE.

Although these methods achieve high performance, lightweight speech enhancement has emerged as an increasingly active area of research. Existing approaches achieve model compression through channel reduction~\cite{Tan2021}, pruning~\cite{Koh2025}, or efficient operations, such as depthwise separable convolutions \cite{Schrter2021DeepfilternetAL,schroeter2023deepfilternet3,Schrter2022deepfilternet2}. 
However, aggressive parameter reduction can significantly degrade a model’s representational capacity, and its impact is particularly critical when modeling phase information~\cite{PALIWAL2011465}. As phase accuracy directly affects perceptual quality, phase distortions can introduce audible artifacts such as musical noise even when the magnitude is accurately reconstructed, ultimately limiting performance on metrics such as the Perceptual Evaluation of Speech Quality (PESQ)~\cite{PESQ}.

\begin{figure}[t]
    \centering
    \includegraphics[width=\columnwidth]{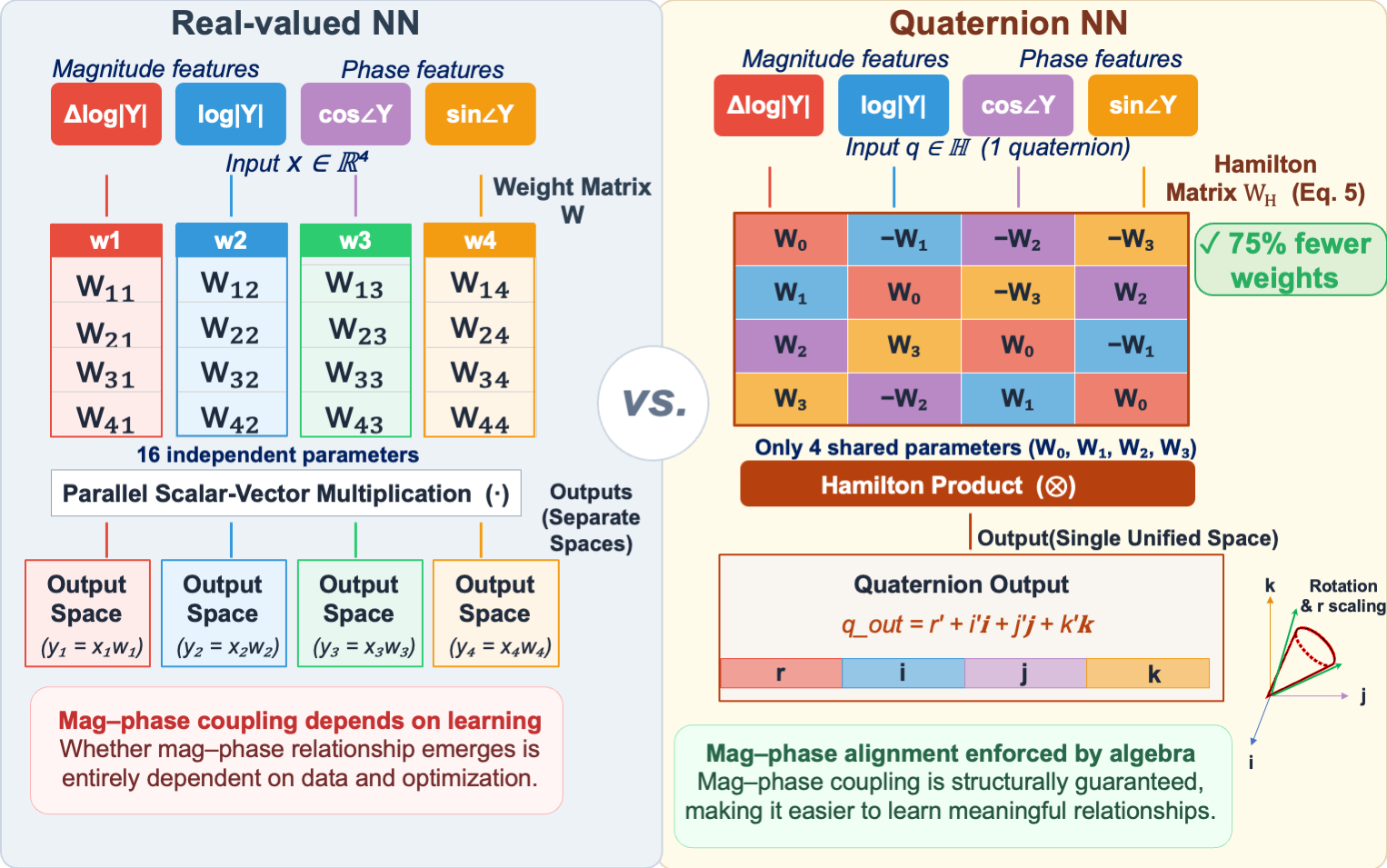}
    \caption{Feature learning in real-valued vs.\ quaternion layers. A real-valued layer considers magnitude and phase independently, whereas a quaternion layer couples them through the Hamilton product as an inductive bias.}
    \label{fig:NN}
\end{figure}

To overcome this trade-off, we focus on Quaternion Neural Networks (QNNs), which leverage Hamilton’s quaternion algebra~\cite{Hamilton1847ONQO}. In a QNN, the Hamilton product performs linear transformations through structured weight sharing by reusing four real-valued sub-matrices across all components, reducing the number of parameters by a factor of four compared with an equivalent real-valued layer~\cite{QuaternionBP,Zhu_2018_ECCV,grant2025,mukhopadhyay2024transformers,yamauchi2025}. Unlike standard real-valued networks that treat each channel independently, this weight sharing imposes a rotation-like coupling among components, enabling joint encoding of the magnitude and phase as a unified quaternion representation~\cite{Parcollet2018cnn,parcollet2018rnn} (Figure~\ref{fig:NN}). This structural inductive bias mitigates the capacity loss inherent to lightweight models.

In this study, we propose a Quaternion Conformer GAN (QC-GAN) that integrates the advantages of quaternions into a modern Conformer architecture. The Conformer captures local spectral patterns through convolution and global temporal dependencies via self-attention; by reformulating these operations in quaternion algebra, the QC-GAN models these features while jointly preserving the magnitude and phase information with significantly fewer parameters. Furthermore, we adopted the MetricGAN training framework, employing a discriminator to approximate perceptual evaluation metrics, such as PESQ, thereby further maximizing the perceptual quality of the enhanced speech. The contributions of this study are summarized as follows:
\begin{itemize}[nosep, leftmargin=*]
\item \textbf{Novel Framework}: To our knowledge, this is the first application of quaternion neural networks to single-channel speech enhancement. Furthermore, we propose the Quaternion Conformer (QC) architecture and demonstrate its effectiveness for achieving this task.
 
\item \textbf{Parameter Efficiency}: QC-GAN achieves higher parameter efficiency than conventional models, with a PESQ of 3.48 using only 0.89M parameters on the VoiceBank+DEMAND dataset, comparable to SoTA models at less than half their size. An ultra-compact variant with only 35K parameters achieves a PESQ of 3.23, further demonstrating the scalability of the proposed framework.
 
\item \textbf{Quaternion Advantage in Phase Preservation}: Ablation studies comparing QC-GAN with an approach in which all quaternion layers are replaced with real-valued layers show that the quaternion representations achieve lower phase errors, demonstrating the inherent advantage of quaternion algebra for phase preservation in speech enhancement.
\end{itemize}

\section{Related Work}
 
\subsection{Speech-Enhancement Models}
Deep learning-based speech enhancement (SE) has evolved from early convolutional neural network (CNN)~\cite{Se2017} and recurrent neural network (RNN) architectures~\cite{Weninger2015} to more sophisticated Transformer and Conformer models. In particular, the Conformer excels at simultaneously capturing local and global dependencies, underpinning SoTA models such as MP-SENet~\cite{Lu_2023} and CMGAN~\cite{Cao2022}. Meanwhile, lightweight SE is an active research area for reducing the model size while preserving performance. For instance, LiSenNet~\cite{Yan2025} combines sub-band processing with a dual-path recurrent module, whereas LSENet~\cite{Koh2025} employs dilated convolutions and attention mechanisms, achieving PESQ scores above 3.0 using extremely compact models with fewer than 0.1M parameters. However, such parameter reduction strategies often impair the modeling of complex spectral structures and phase information when the parameter counts become extremely low, creating a performance gap compared with full-scale models.
 
\subsection{Quaternion Neural Networks in the Audio Domain}
Quaternion Neural Networks (QNNs) embed a four-component input into a four-dimensional hypercomplex space, where the Hamilton product processes it as a single coupled entity, typically achieving a fourfold parameter reduction. In speech and audio processing, QNNs have been explored mainly for \emph{discriminative} tasks: Parcollet et al.~\cite{Parcollet2018cnn,parcollet2018rnn} assign Mel filter-bank energies and their temporal derivatives to the quaternion axes for automatic speech recognition (ASR), with related formulations extended to multi-channel distant ASR~\cite{qiu2020quaternion} and 3D sound source localization~\cite{Celsi2020}. These studies predict labels or spatial estimates rather than \emph{reconstructing} a clean waveform, and their quaternion axes encode hand-crafted acoustic descriptors rather than the spectrogram directly. To our knowledge, no prior study has applied QNNs to single-channel speech enhancement, where magnitude and phase must be jointly estimated. We address this gap by assigning the short-time Fourier transform (STFT) magnitude and phase directly to the quaternion axes and reformulating the Conformer— encompassing not only convolutional and recurrent layers but also its architecture as a whole—within hypercomplex algebra, positioning phase reconstruction as the locus where the Hamilton-product coupling is expected to fully exploit its advantage.
 
\subsection{Metric-based Optimization}
Conventional SE models are typically trained using objective functions such as mean squared error (MSE) on spectrograms. However, these loss functions do not necessarily correlate with human auditory perception or evaluation metrics such as PESQ~\cite{PESQ} and Short-Time Objective Intelligibility (STOI)~\cite{STOI}. To bridge this gap, the MetricGAN~\cite{Fu2019MetricGAN} framework and its successors, MetricGAN+~\cite{Fu2021MetricGANPlus} and CMGAN~\cite{Cao2022}, trained a discriminator to estimate evaluation metrics, guiding the generator toward optimizing perceptual quality. The proposed QC-GAN follows this paradigm. However, unlike CMGAN, which relies on a real-valued Conformer, we employ a Quaternion Conformer as the generator. Quaternion operations enable the generator to model the magnitude and phase with substantially fewer parameters, whereas MetricGAN-based training further maximizes perceptual quality.

\section{Proposed Method}
 
This section describes the proposed QC-GAN architecture. Section~\ref{subsec:quat_algebra} introduces the fundamentals of quaternion algebra, and Section~\ref{subsec:quat_layers} defines the quaternion neural network layers used as building blocks. Sections~\ref{subsec:architecture} and~\ref{subsec:loss} present the overall QC-GAN architecture and its loss functions, respectively. Finally, Section~\ref{subsec:quat_feature} describes the quaternion feature representation used as the model input. 
 
\begin{figure*}[t]
    \centering
    \includegraphics[width=0.88\textwidth]{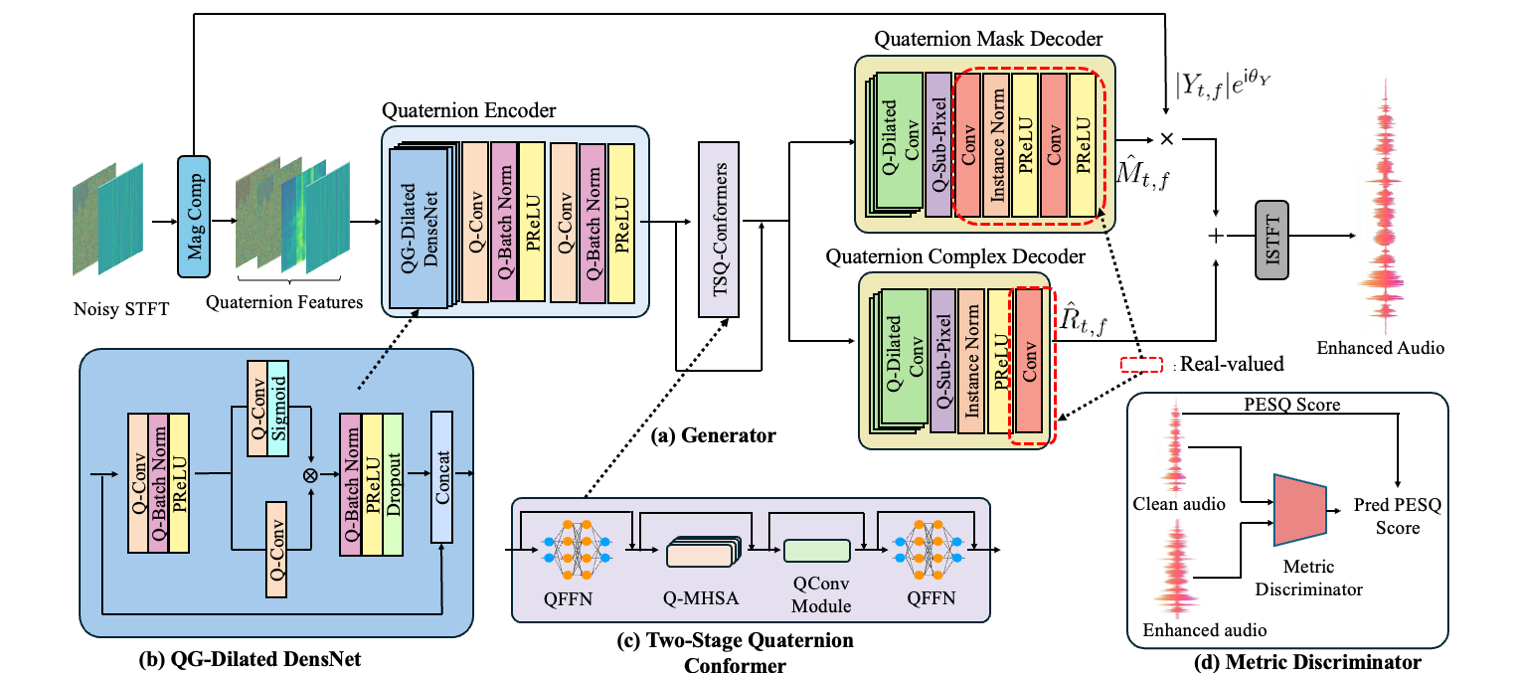}
    \caption{Overall architecture of the proposed QC-GAN, comprising a Quaternion Encoder, a QG-Dilated DenseNet encoder (b), a two-stage Quaternion Conformer bottleneck (c), a dual-branch decoder (mask and complex residual), and a metric discriminator (d). The overall generator architecture is shown in (a).}
 
    \label{fig:model_archtecture}
\end{figure*}

\subsection{Quaternion Algebra}
\label{subsec:quat_algebra}
 
A quaternion $q \in \mathbb{H}$ is a hypercomplex number with one real $q_0$ and three imaginary components $q_1, q_2, q_3$:
\begin{equation}
  q = q_0 + q_1 {\sf i} + q_2 {\sf j} + q_3 {\sf k},
\end{equation}
where $\mathbb{H}$ denotes the set of quaternions. The imaginary units ${\sf i}$, ${\sf j}$, and ${\sf k}$ obey the following fundamental relations:
\begin{equation}
  {\sf i}^2 = {\sf j}^2 = {\sf k}^2 
  = {\sf ijk} = -1.
\end{equation}
The quaternion conjugate and its squared norm are defined as
\begin{equation}
  q^* = q_0 - q_1 {\sf i} - q_2 {\sf j} - q_3 {\sf k}, 
  \qquad 
  \lVert q \rVert^2 = q \otimes q^*,
\end{equation}
where $\otimes$ denotes the Hamilton product. The Hamilton product of two quaternions $p, q \in \mathbb{H}$ is given by
\begin{align}
\label{eq:hamilton_product}
p \otimes q
 &=       (p_0 q_0 - p_1 q_1 - p_2 q_2 - p_3 q_3) \nonumber\\
 &\quad + (p_0 q_1 + p_1 q_0 + p_2 q_3 - p_3 q_2){\sf i} \nonumber\\
 &\quad + (p_0 q_2 - p_1 q_3 + p_2 q_0 + p_3 q_1){\sf j} \nonumber\\
 &\quad + (p_0 q_3 + p_1 q_2 - p_2 q_1 + p_3 q_0){\sf k}.
\end{align}

\subsection{Quaternion Neural Networks Layer}   
\label{subsec:quat_layers}
 
Building upon the Hamilton product defined in Section~\ref{subsec:quat_algebra}, we define the proposed QC-GAN’s foundational elements: the Quaternion fully-connected layer (QFC) and the Quaternion Convolutional layer (QConv).
 
\subsubsection{Quaternion Fully-Connected Layer}
\label{sec:QFC}
 
Let $\mathbf{x} = \mathbf{x}_0 + \mathbf{x}_1 {\sf i} + \mathbf{x}_2 {\sf j} + \mathbf{x}_3 {\sf k} \in \mathbb{H}^{N_{in}}$ be the input vector and $\mathbf{W} = \mathbf{W}_0 + \mathbf{W}_1 {\sf i} + \mathbf{W}_2 {\sf j} + \mathbf{W}_3 {\sf k} \in \mathbb{H}^{N_{out} \times N_{in}}$ be the quaternion weight matrix, where $N_{in}$ and $N_{out}$ denote the input and output dimensions in the quaternion domain, respectively, and $\mathbf{W}_{\{0,1,2,3\}}$ are real-valued matrices~\cite{Gaudet2018}. The linear transformation $\mathbf{y} = \mathbf{W} \otimes \mathbf{x}$ can be formulated as a real-valued matrix-vector multiplication using the Hamilton matrix $\mathbf{H}_W$:
\begin{equation}
    \renewcommand{\arraystretch}{1.3}
    \begin{bmatrix}
        \mathbf{y}_0 \\ \mathbf{y}_1 \\ \mathbf{y}_2 \\ \mathbf{y}_3
    \end{bmatrix}
    =
    \underbrace{
    \begin{bmatrix}
         \mathbf{W}_0 & -\mathbf{W}_1 & -\mathbf{W}_2 & -\mathbf{W}_3 \\
         \mathbf{W}_1 &  \mathbf{W}_0 & -\mathbf{W}_3 &  \mathbf{W}_2 \\
         \mathbf{W}_2 &  \mathbf{W}_3 &  \mathbf{W}_0 & -\mathbf{W}_1 \\
         \mathbf{W}_3 & -\mathbf{W}_2 &  \mathbf{W}_1 &  \mathbf{W}_0
    \end{bmatrix}
    }_{\makebox[0pt]{\raisebox{-8pt}{$\mathbf{H}_W$}}}
    \begin{bmatrix}
        \mathbf{x}_0 \\ \mathbf{x}_1 \\ \mathbf{x}_2 \\ \mathbf{x}_3
    \end{bmatrix}.
    \label{eq:qfc_matrix}
\end{equation}

Here, $\mathbf{H}_W \in \mathbb{R}^{4N_{out} \times 4N_{in}}$ denotes the Hamilton matrix constructed from the components of $\mathbf{W}$. As shown in Eq.~\eqref{eq:qfc_matrix}, the sub-matrices $\mathbf{W}_{\{0,1,2,3\}}$ are shared across different axes. Consequently, a QFC layer requires only $4 \times N_{in} \times N_{out}$ parameters, achieving a 75\% reduction compared with a real-valued layer with an equivalent input--output dimensionality of ($4N_{in} \times 4N_{out}$).

\subsubsection{Quaternion Convolutional Layer}
Quaternion convolution extends the Hamilton product to the convolution operation~\cite{Gaudet2018}. Let $\mathcal{X} \in \mathbb{H}^{T \times F \times C_{in}}$ be the input feature map and $\mathcal{K} \in \mathbb{H}^{K_t \times K_f \times C_{in} \times C_{out}}$ be the quaternion kernel, defined as $\mathcal{K} = \mathcal{K}_0 + \mathcal{K}_1 {\sf i} + \mathcal{K}_2 {\sf j} + \mathcal{K}_3 {\sf k}$, where $T$ is the number of time frames, $F$ is the number of frequency bins, $C_{in}$ and $C_{out}$ are the numbers of input and output quaternion channels, respectively, and $K_t \times K_f$ is the kernel size. The output feature map $\mathcal{Y} = \mathcal{K} \ast \mathcal{X}$ is calculated by convolving the four real-valued component feature maps according to the Hamilton product rule (Eq.~\eqref{eq:hamilton_product}):
 
\begin{equation}
    \begin{aligned}
        \mathcal{Y}_0 &= \mathcal{K}_0 \ast \mathcal{X}_0 - \mathcal{K}_1 \ast \mathcal{X}_1 - \mathcal{K}_2 \ast \mathcal{X}_2 - \mathcal{K}_3 \ast \mathcal{X}_3, \\
        \mathcal{Y}_1 &= \mathcal{K}_0 \ast \mathcal{X}_1 + \mathcal{K}_1 \ast \mathcal{X}_0 + \mathcal{K}_2 \ast \mathcal{X}_3 - \mathcal{K}_3 \ast \mathcal{X}_2, \\
        \mathcal{Y}_2 &= \mathcal{K}_0 \ast \mathcal{X}_2 - \mathcal{K}_1 \ast \mathcal{X}_3 + \mathcal{K}_2 \ast \mathcal{X}_0 + \mathcal{K}_3 \ast \mathcal{X}_1, \\
        \mathcal{Y}_3 &= \mathcal{K}_0 \ast \mathcal{X}_3 + \mathcal{K}_1 \ast \mathcal{X}_2 - \mathcal{K}_2 \ast \mathcal{X}_1 + \mathcal{K}_3 \ast \mathcal{X}_0,
    \end{aligned}
    \label{eq:qconv_ops}
\end{equation}
where $\ast$ denotes a standard real-valued convolution.

\subsubsection{Quaternion Multi-Head Self-Attention}
Standard self-attention mechanisms rely on real-valued dot products and treat feature channels independently. In contrast, we employ the Quaternion Multi-Head Self-Attention (Q-MHSA) proposed by Tay et al.~\cite{Tay2019}, which captures latent structural dependencies between quaternion components.
 
Given a quaternion input sequence $\mathbf{X} \in \mathbb{H}^{T \times D}$, where $D$ denotes the quaternion feature dimension, the query, key, and value matrices ($\mathbf{Q}, \mathbf{K}, \mathbf{V}$) are projected using the QFC layers defined in Section~\ref{sec:QFC}:
\begin{equation}
    \mathbf{Q} = \mathbf{W}_q \otimes \mathbf{X}, \quad \mathbf{K} = \mathbf{W}_k \otimes \mathbf{X}, \quad \mathbf{V} = \mathbf{W}_v \otimes \mathbf{X},
\end{equation}
where $\mathbf{W}_q, \mathbf{W}_k, \mathbf{W}_v \in \mathbb{H}^{D \times D}$ are quaternion weight matrices. 
 
The attention scores are computed using the Hamilton product of the queries and keys, followed by a scaling factor and a softmax function:
\begin{equation}
    \mathbf{A}_{\alpha} = \text{ComponentSoftmax}\left(\frac{\mathbf{Q} \otimes \mathbf{K}^\top}{\sqrt{d_k}}\right), \quad \alpha \in \{0,1,2,3\}
    \label{eq:q_attention_score}
\end{equation}
 
where $d_k = D / h$ denotes the dimension of each head, and $h$ is the number of attention heads. 
Crucially, as formulated by Tay et al.~\cite{Tay2019}, the softmax function in Eq.~\eqref{eq:q_attention_score} is applied component-wise to each of the real and imaginary parts (indexed by $0, 1, 2, 3$) independently (ComponentSoftmax).
Finally, the output of the attention head is computed as $\mathbf{Y} = \mathbf{A}_{\alpha} \mathbf{V}$, with RMSNorm applied to queries and keys for scale stability. The outputs from all heads are concatenated and projected back to the original dimension via a QFC layer.

\subsection{QC-GAN Architecture}
\label{subsec:architecture}
QC-GAN, inspired by the CMGAN architecture~\cite{Cao2022}, reformulates its components in the quaternion domain (Figure~\ref{fig:model_archtecture} (a)). The generator comprises a quaternion encoder, a two-stage Quaternion Conformer bottleneck (c), and a dual-branch decoder that simultaneously estimates the magnitude masks and complex residuals. A metric discriminator (d) approximates perceptual scores, such as PESQ, to further maximize the enhancement quality.

\subsubsection{Encoder}
The encoder projects the input quaternion features $\mathbf{X} \in \mathbb{H}^{B \times 1 \times T \times F}$, where $B$ is the batch size, $T$ is the number of time frames, $F$ is the number of frequency bins, and $1$ denotes a single quaternion channel (equivalent to four real-valued channels), into a latent representation using the Quaternion Gated (QG)-Dilated DenseNet (Figure~\ref{fig:model_archtecture}(b)). Dense connections reuse features from earlier layers, enabling effective feature extraction with fewer parameters \cite{PandeyDensenet}. Each dense block applies quaternion dilated convolutions with dilation rates of $1, 2, 4$ and $8$ to capture multi-scale spectral patterns. A gating mechanism multiplies the convolution output element-wise with a parallel sigmoid branch~\cite{vandenoord16_ssw}, suppressing the noise-related activations while retaining informative features. We utilized Quaternion Batch Normalization (Q-Batch Norm) based on quaternion variance statistics, as proposed in~\cite{q-batch}.
The encoder concludes by downsampling the frequency dimension ($F/2$), yielding a latent tensor of shape $[B, C, T, F/2]$ for efficient processing in the subsequent bottleneck.

\subsubsection{Bottleneck}
The bottleneck employs a Two-Stage Quaternion Conformer (TSQ-Conformer, Figure~\ref{fig:model_archtecture}(c)), which stacks $N$ Q-Conformer blocks, each of which applies self-attention, first along the time dimension and then along the frequency dimension~\cite{Lu_2023}. This two-stage design enables the model to capture global dependencies across the entire spectrogram while preserving its quaternion structure. A residual connection is applied around the bottleneck module to facilitate the gradient flow.

\subsubsection{Decoder}
The decoder employs a dual-branch architecture inspired by CMGAN~\cite{Cao2022} to reconstruct high-resolution T--F features.
It comprises two parallel paths: a mask branch that estimates a real-valued magnitude mask $\hat{M}_{t,f}$ and a complex branch that predicts complex residuals $\hat{R}_{t,f}$ to correct the phase details.
Each branch first applies a dilated quaternion convolutional block (with dilation rates $d=\{1, 2, 4, 8\}$) to capture wide-context dependencies, followed by a Quaternion Sub-Pixel Convolution ($r=2$) that upsamples the frequency dimension ($F/2 \to F$) while preventing checkerboard artifacts.
The mask branch concludes with channel-wise, learnable PReLU activations~\cite{prelu}. Both branches project the quaternion features into the real domain through a final real-valued convolution.
Finally, the outputs are integrated to synthesize the enhanced complex spectrogram bin $\hat{S}_{t,f}$.
 
Let $Y_{t,f} \in \mathbb{C}$ denote the noisy STFT coefficient. It can be expressed in polar form as
$Y_{t,f} = |Y_{t,f}|e^{\mathrm{i}\theta_{Y_{t,f}}}$, where $\theta_{Y_{t,f}}=\angle Y_{t,f}$ denotes the noisy phase.
We first compute the masked magnitude and its phase-carrying complex spectrum as
\begin{align}
    \tilde{S}_{t,f} = \hat{M}_{t,f}\,|Y_{t,f}|\,e^{\mathrm{i}\theta_{Y_{t,f}}}.
\end{align}
The enhanced complex spectrogram is then reconstructed by adding the predicted complex residual:
\begin{equation}
    \hat{S}_{t,f} = \tilde{S}_{t,f} + \hat{R}_{t,f}.
\end{equation}
The final time-domain waveform is obtained by applying the Inverse STFT (ISTFT) to $\hat{S}$.

\subsection{Loss Functions}
\label{subsec:loss}
To enable high-fidelity speech reconstruction and optimize perceptual quality, we trained QC-GAN using a multi-task objective function comprising a generator loss $\mathcal{L}_{G}$ and a discriminator loss $\mathcal{L}_{D}$.
 
\subsubsection{Generator Loss}
The total generator loss is defined as a weighted sum of five components:
\begin{equation}
    \mathcal{L}_{G} = 0.1\mathcal{L}_{\text{RI}} + 0.9\mathcal{L}_{\text{Mag}} + 0.2\mathcal{L}_{\text{Time}} + 0.05\mathcal{L}_{\text{PESQ}} + 0.05\mathcal{L}_{\text{GAN}}.
\end{equation}
The loss weights were empirically determined based on preliminary experiments.
The individual loss terms are defined as
\begin{align}
    &\mathcal{L}_{\text{RI}} = \mathbb{E} \left[ \lVert \hat{S}_r - S_r \rVert^2 \right] + \mathbb{E} \left[ \lVert \hat{S}_i - S_i \rVert^2 \right], \\
    &\mathcal{L}_{\text{Mag}} = \mathbb{E} \left[ \lVert \hat{A} - A \rVert^2 \right], \\
    &\mathcal{L}_{\text{Time}} = \mathbb{E} \left[ \lVert \hat{y} - y \rVert_1 \right], \\
    &\mathcal{L}_{\text{PESQ}} = \mathbb{E} \left[ \Psi^{(s)}(\mathbf{l}, \hat{\mathbf{l}}) + \Psi^{(a)}(\mathbf{l}, \hat{\mathbf{l}}) \right], \\
    &\mathcal{L}_{\text{GAN}} = \mathbb{E} \left[ \lVert D(A, \hat{A}) - 1 \rVert^2 \right],
\end{align}
where $S$ and $\hat{S}$ respectively denote the clean and enhanced complex spectrograms, with subscripts $r$ and $i$ indicating their real and imaginary parts;
$A=|S|$ and $\hat{A}=|\hat{S}|$ are the corresponding magnitude spectrograms;
$y$ and $\hat{y}$ are the clean and enhanced time-domain waveforms;
$\Psi^{(s)}$ and $\Psi^{(a)}$ are the symmetrical and asymmetrical disturbance terms computed from loudness spectra $\mathbf{l}$ and $\hat{\mathbf{l}}$ derived from $y$ and $\hat{y}$~\cite{Martin2018};
and $D$ is the discriminator. 
The differentiable PESQ loss provides a stable perceptual training signal, whereas the MetricGAN discriminator offers adaptive guidance from the data distribution, and their combination facilitates perceptual quality optimization.

\subsubsection{Discriminator Loss}
The discriminator $D$ acts as a learned metric approximator. It takes a pair of magnitude spectrograms (reference and target) as input. The discriminator loss $\mathcal{L}_{D}$ minimizes the prediction error for both clean and enhanced speech as follows:
\begin{equation}
    \mathcal{L}_{D} =
    \mathbb{E} \left[ \left\lVert D(A, A) - 1 \right\rVert^2 \right]
    + \mathbb{E} \left[ \left\lVert D(A, \hat{A}) - Q_{\text{PESQ}} \right\rVert^2 \right],
\end{equation}
where $D$ denotes the discriminator and $Q_{\text{PESQ}}$ the normalized PESQ score in the range $[0,1]$.

\subsection{Quaternion Feature}
\label{subsec:quat_feature}
To capture the spectral dynamics and phase relations effectively, we constructed a four-channel quaternion input
$\mathbf{X}_{\text{in}} \in \mathbb{H}^{B \times 1 \times T \times F}$ from the noisy speech STFT coefficients $Y$.
Let $|Y|_\epsilon$ denote the magnitude spectrum clamped by a small constant $\epsilon$ (e.g., $10^{-8}$) for numerical stability, defined as
\begin{equation}
    |Y_{t,f}|_\epsilon = \max(|Y_{t,f}|, \epsilon).
\end{equation}
The quaternion feature is represented as four real-valued channels of shape $[B,4,T,F]$, and is defined as
\begin{align}
\label{eq:q-features}
    \mathbf{X}_{\text{in}} = \Delta \log(|Y|_\epsilon) + \log(|Y|_\epsilon)\,\mathsf{i}
    + \frac{\operatorname{Re}(Y)}{|Y|_\epsilon}\,\mathsf{j} + \frac{\operatorname{Im}(Y)}{|Y|_\epsilon}\,\mathsf{k}.
\end{align}
The real part is assigned the first-order temporal difference of the log-magnitude, $\Delta\log(|Y|_\epsilon)$, to capture the spectral dynamics. Concretely,
\begin{equation}
    \Delta\log(|Y|_\epsilon)_{t,f} =
    \begin{cases}
        0, & t=0,\\
        \log(|Y_{t,f}|_\epsilon)-\log(|Y_{t-1,f}|_\epsilon), & t>0,
    \end{cases}
\end{equation}
That is, the difference is computed along the time axis, and the first frame is padded with zeros.
The first imaginary component (associated with unit $\mathsf{i}$) encodes the log-magnitude $\log(|Y|_\epsilon)$, which represents the static spectral envelope of the signal.
The remaining imaginary components (associated with units $\mathsf{j}$ and $\mathsf{k}$) correspond to the cosine and sine of the phase normalized by the clamped magnitude, respectively, thereby preserving the circular phase structure within the quaternion domain.
Following CMGAN and MP-SENet~\cite{Cao2022,Lu_2023}, we applied power-law compression ($\beta=0.3$) to the complex STFT coefficients and computed the spectral-domain losses (and the discriminator) in the compressed domain. The predicted spectra were power-uncompressed before the ISTFT reconstruction.

\section{Experiment}

\subsection{Dataset}
\subsubsection{VoiceBank+DEMAND}
We evaluated QC-GAN using the VoiceBank+DEMAND dataset~\cite{valentinibotinhao16_ssw}\footnote{https://huggingface.co/datasets/JacobLinCool/VoiceBank-DEMAND-16k}, which comprises 28 speakers (11,572 utterances) for training and 2 speakers (824 utterances) for testing. Clean speech from the Voice Bank corpus~\cite{veaux2013voicebank} was mixed with noise from the DEMAND database~\cite{thiemann2013demand} and artificial sources at SNRs of 0--15~dB for training, and with unseen noise types at SNRs of 2.5--17.5~dB for testing. Using this dataset, we first compared QC-GAN with recent SoTA models to demonstrate its performance. Furthermore, to highlight the parameter efficiency of our framework, we conducted a focused comparison with ultra-compact models to verify the superiority of our compact variant, QC-GAN (Tiny).

\subsubsection{DNS-Challenge 3}
Although the VoiceBank+DEMAND dataset comprises synthetically mixed noise under controlled conditions, real-world deployment requires robustness to diverse and unpredictable acoustic environments. To evaluate this, we employed the Deep Noise Suppression Challenge 3 (DNS-Challenge 3) dataset~\cite{reddy2021interspeech}\footnote{https://github.com/microsoft/DNS-Challenge}, released as part of an international benchmark for speech enhancement under challenging acoustic conditions, which provides approximately 760 hours of clean speech covering diverse speaking styles and languages, 181 hours of noise spanning approximately 150 classes, and over 118,000 room impulse responses for reverberant simulation. We used the blind test set to verify whether QC-GAN, despite its compact size, can generalize to large-scale and diverse conditions without underfitting.

\subsection{Experimental Setup}
All audio recordings were resampled to 16 kHz. During training, the audio segments were cropped to a fixed length of 2 s (32{,}000 samples). For spectral analysis, we computed the STFT using a Hann window of length 25 ms (400 samples) and hop size 6.25 ms (100 samples).
The models were trained using the AdamW~\cite{loshchilov2018decoupled} optimizer with $\beta_1=0.5, \beta_2=0.999$. 
The initial learning rates were set to $5 \times 10^{-4}$ for the generator and $1 \times 10^{-3}$ for the discriminator, with a decay schedule applied every 30 epochs. Gradient clipping was set to 1.0 to stabilize the training. We trained for 100 epochs on the VoiceBank+DEMAND dataset and 50 epochs on the DNS-Challenge dataset, each trained separately from scratch.

We configured two model variants: QC-GAN (Base) and QC-GAN (Tiny). Both models use four attention heads. The Base model uses a growth rate (output channels per dense block) of 64, 4 TSQ-Conformer layers, and 64 decoder channels (0.89M parameters), whereas the Tiny model reduces these to 16, 1, and 16, respectively (35K parameters).

To comprehensively assess enhancement performance, we employed distinct metrics tailored to each dataset.
For VoiceBank+DEMAND, we reported standard objective metrics: PESQ~\cite{PESQ}, STOI~\cite{STOI}, and composite metrics comprising CSIG (signal distortion), CBAK (background intrusiveness), and COVL (overall quality)~\cite{Hu2008}.
To characterize both parameter efficiency and computational cost, 
we report the number of parameters (Params) and, for the 
ultra-compact models, the number of real-valued MACs.
For DNS-Challenge 3, given the absence of clean references for the blind test set, we used non-intrusive metrics: DNSMOS P.835 scores~\cite{reddy2022dnsmos}---Signal (SIG), Background (BAK), and Overall (OVRL)---along with the ITU-T P.808 MOS~\cite{reddy2021dnsmos} as a perceptual quality proxy.

\subsection{Results}
\subsubsection{VoiceBank+DEMAND}

\begin{table*}[t]
  \caption{Performance comparison with state-of-the-art methods on VoiceBank+DEMAND. Bold indicates the best performance in each metric.}
  \label{tab:SoTA_results}
  \centering
    \small
  \begin{tabular}{l|c|ccccc}
    \toprule
    \textbf{Model} & \textbf{Params (M)} & \textbf{PESQ} & \textbf{STOI} & \textbf{CSIG} & \textbf{CBAK} & \textbf{COVL} \\
    \midrule
    Noisy      & -    & 1.97 & 0.91  & 3.35 & 2.44 & 2.63        \\
    \midrule
    SEGAN~\cite{pascual2017segan}         & 97.47 & 2.16 & 0.92 & 3.48 & 2.94 & 2.80  \\
    MetricGAN+~\cite{Fu2021MetricGANPlus} & - & 3.15 & - & 4.14 & 3.16 & 3.64 \\
    DPT-FSNet~\cite{Dang2021DPTFSNetDT}   & \textbf{0.88} & 3.33 & 0.96 & 4.58 & 3.72 & 4.00 \\
    CMGAN~\cite{Cao2022}                  & 1.83 & 3.41 & \textbf{0.96} & 4.63 & 3.94 & 4.12 \\
    MP-SENet~\cite{Lu_2023}               & 2.05 & 3.50 & \textbf{0.96} & 4.73 & \textbf{3.95} & 4.22 \\
    SE-Mamba~\cite{chao2024}          & 2.26 & \textbf{3.55} & \textbf{0.96} & \textbf{4.77} & \textbf{3.95} & \textbf{4.26}\\
    \midrule
    \textbf{QC-GAN (Base)} & 0.89 & 3.48 & 0.95 & 4.60 & 3.65 & 4.10 \\
    \bottomrule
  \end{tabular}%
\end{table*}
 
\begin{table}[t]
\caption{Comparison with ultra-compact models on VoiceBank+DEMAND. MACs are real-valued. For QC-GAN, the value in parentheses is the qMAC count of the parametrized layers (0.22\,G; $1$ qMAC $=16$ real MACs); $^{\dagger}$its real-MAC value additionally includes attention.}
 
  \label{tab:lightweight_results}
  \centering
  \setlength{\tabcolsep}{3.5pt}
  \resizebox{\columnwidth}{!}{
  \begin{tabular}{l|c|c|cc}
    \toprule
    \textbf{Model} & \textbf{Params (K)} & \textbf{MACs(G)} & \textbf{PESQ} & \textbf{STOI} \\
    \midrule
    Noisy & -- & -- & 1.97 & 0.91 \\
    \midrule
    RNNoise~\cite{valin2018rnnoise} & 60 & 0.04 & 2.33 & 0.92 \\
    CCFNet+ (Lite)~\cite{dang2023lightweight} & 160 & 0.39 & 2.94 & - \\
    FSPEN~\cite{yang2024fspen} & 79 & 0.09 & 2.97 & 0.94 \\
    LiSenNet~\cite{Yan2025} & 37 & 0.06 & 3.07 & 0.94 \\
    LSENet~\cite{Koh2025} & 39 & 0.24 & 3.12 & \textbf{0.95}\\
    \midrule
    \textbf{QC-GAN (Tiny)} & \textbf{35} & 3.75 (0.22)$^{\dagger}$ & \textbf{3.23} & 0.94 \\
    \bottomrule
  \end{tabular}
  }
\end{table}

The Base model is compared with recent state-of-the-art methods (CMGAN~\cite{Cao2022}, MP-SENet~\cite{Lu_2023}, SE-Mamba~\cite{chao2024}), DPT-FSNet~\cite{Dang2021DPTFSNetDT} (with a comparable parameter count), and MetricGAN+~\cite{Fu2021MetricGANPlus} (a representative metric-based approach). 
The Tiny model is compared with (LSENet~\cite{Koh2025}, LiSenNet~\cite{Yan2025}, FSPEN~\cite{yang2024fspen}, CCFNet+~\cite{dang2023lightweight}) and RNNoise~\cite{valin2018rnnoise} as a baseline. The results are given in Tables~\ref{tab:SoTA_results} and \ref{tab:lightweight_results}.
 
Table~\ref{tab:SoTA_results} compares QC-GAN (Base) with SoTA models. With only 0.89M parameters, comparable to DPT-FSNet~\cite{Dang2021DPTFSNetDT}, QC-GAN achieves a PESQ of 3.48, surpassing CMGAN~\cite{Cao2022} (PESQ 3.41) and approaching MP-SENet~\cite{Lu_2023} (PESQ 3.50) and SE-Mamba~\cite{chao2024} (PESQ 3.55), which require approximately 2--2.5$\times$ more parameters.
 
Table~\ref{tab:lightweight_results} focuses on the ultra-compact regime. QC-GAN (Tiny) achieves a PESQ of 3.23 with only 35K parameters, outperforming LiSenNet~\cite{Yan2025} (PESQ 3.07, 37K), LSENet~\cite{Koh2025} (PESQ 3.12, 39K), and RNNoise~\cite{valin2018rnnoise} (PESQ 2.33, 60K) at a comparable or smaller model size. An STOI of 0.94 confirmed that speech intelligibility was preserved. These results demonstrate that QC-GAN effectively captures the speech structures even under extreme parameter constraints.

\subsubsection{DNS-Challenge}
\begin{table}[t]
  \caption{Performance comparison on the DNS-Challenge 3 blind test set using DNSMOS metrics. All models were trained from scratch on the DNS-Challenge 3 training data.}
 
  \label{tab:dns_results}
  \centering
  \resizebox{\columnwidth}{!}{
  \begin{tabular}{l|c|cccc}
    \toprule
    \textbf{Model} & \textbf{Params} & \textbf{OVRL} & \textbf{SIG} & \textbf{BAK} & \textbf{P808\_MOS} \\
    \midrule
    Noisy & -- & 2.11 & 2.89 & 2.34 & 2.92 \\
    \midrule
    NSNet2~\cite{braun2020dataaugmentationlossnormalization} & 2.7M & 2.31 & 2.89 & 2.85 & 3.01 \\
    DCCRN~\cite{Hu2020DCCRNDC} & 3.7M & 2.54 & 2.98 & 3.43 & 3.26 \\
    CMGAN~\cite{Cao2022}& 1.83M & 2.66 & \textbf{3.08} & 3.59 & 3.12 \\
    \midrule
    QC-GAN (Tiny) & \textbf{35K} & 2.56 & 2.95 & 3.54 & 3.23 \\
    QC-GAN (Base) & 0.89M & \textbf{2.73} & 3.07 & \textbf{3.79} & \textbf{3.37} \\
    \bottomrule
  \end{tabular}
  }
\end{table}
 
To validate the generalization to real-world conditions, we evaluated QC-GAN on the DNS-Challenge 3 blind test set (Table~\ref{tab:dns_results}). We compared the proposed model with NSNet2~\cite{braun2020dataaugmentationlossnormalization}, the official baseline of the DNS-Challenge, DCCRN~\cite{Hu2020DCCRNDC}, a representative model from previous challenges, and CMGAN~\cite{Cao2022}, a real-valued Conformer-based architecture that serves as the closest baseline. For fair comparison, all models were trained from scratch using the DNS-Challenge 3 training data.
 
Among all the compared methods, QC-GAN (Base) achieved the highest scores in OVRL (2.73), BAK (3.79), and P.808 MOS (3.37), outperforming CMGAN despite using approximately half of its parameters. QC-GAN (Tiny), despite having only 35K parameters, surpassed NSNet2 (2.7M) across all metrics and achieved performance comparable to DCCRN (3.7M). These results confirm that QC-GAN generalizes effectively, even under complex real-world noise conditions, without relying on excessive model capacity.

\section{Ablation Study}
 
\begin{table}[t]
  \caption{Ablation study: QC-GAN (35K params) vs.\ Real-NN (32K, 140K params) with comparable and matched-capacity model sizes. TS-Conformer denotes the real-valued Two-Stage Conformer.}
  \label{tab:ablation_quaternion}
  \centering
  \resizebox{\columnwidth}{!}{
    \begin{tabular}{l|ccccc}
        \toprule
        \textbf{Model} & \textbf{PESQ} & \textbf{STOI} & \textbf{CSIG} & \textbf{CBAK} & \textbf{COVL} \\
        \midrule
        Real-NN (32K)           & 3.12 & 0.94 & 4.29 & 3.30 & 3.73 \\
        \textbf{Real-NN (140K)} & \textbf{3.29} & \textbf{0.94} & \textbf{4.45} & \textbf{3.48} & \textbf{3.90} \\
        \quad w/o Discriminator & 3.32 & 0.94 & 4.45 & 3.40 & 3.76 \\
        \quad w/o TS-Conformer  & 3.10 & 0.93 & 4.03 & 3.25 & 3.57 \\
        \midrule
        \textbf{QC-GAN (Tiny Full)} & \textbf{3.23} & \textbf{0.94} & \textbf{4.33} & \textbf{3.36} & \textbf{3.79} \\
        \quad w/o Discriminator & 3.14 & 0.94 & 4.33 & 3.40 & 3.76 \\
        \quad w/o TSQ-Conformer & 2.99 & 0.93 & 3.91 & 3.19 & 3.44 \\
        \bottomrule
    \end{tabular}
  }
\end{table}
 
\begin{figure}[t]
    \centering
    \includegraphics[width=0.85\columnwidth]{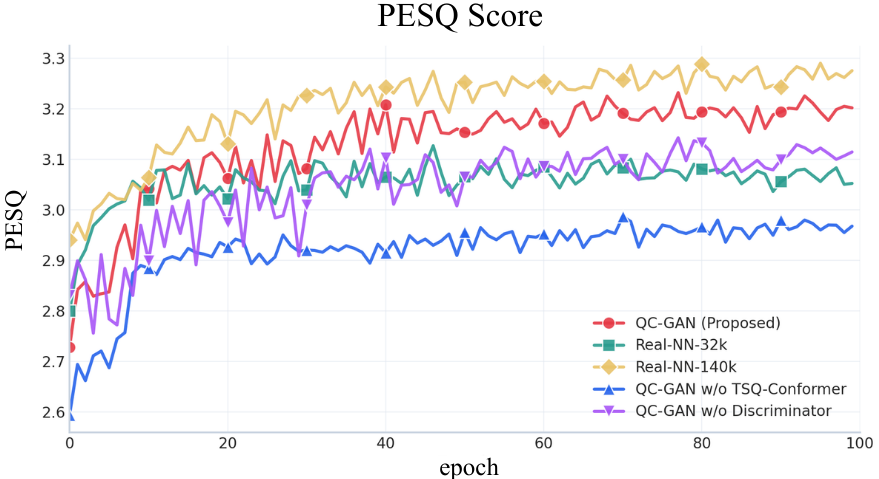}
    \caption{PESQ curves over 100 training epochs for QC-GAN, Real-NN (32K and 140K), and QC-GAN ablated variants (w/o Discriminator and w/o TSQ-Conformer).}
    \label{fig:pesq_curve}
\end{figure}
 
\begin{figure}[t]
    \centering
    \includegraphics[width=0.90\columnwidth]{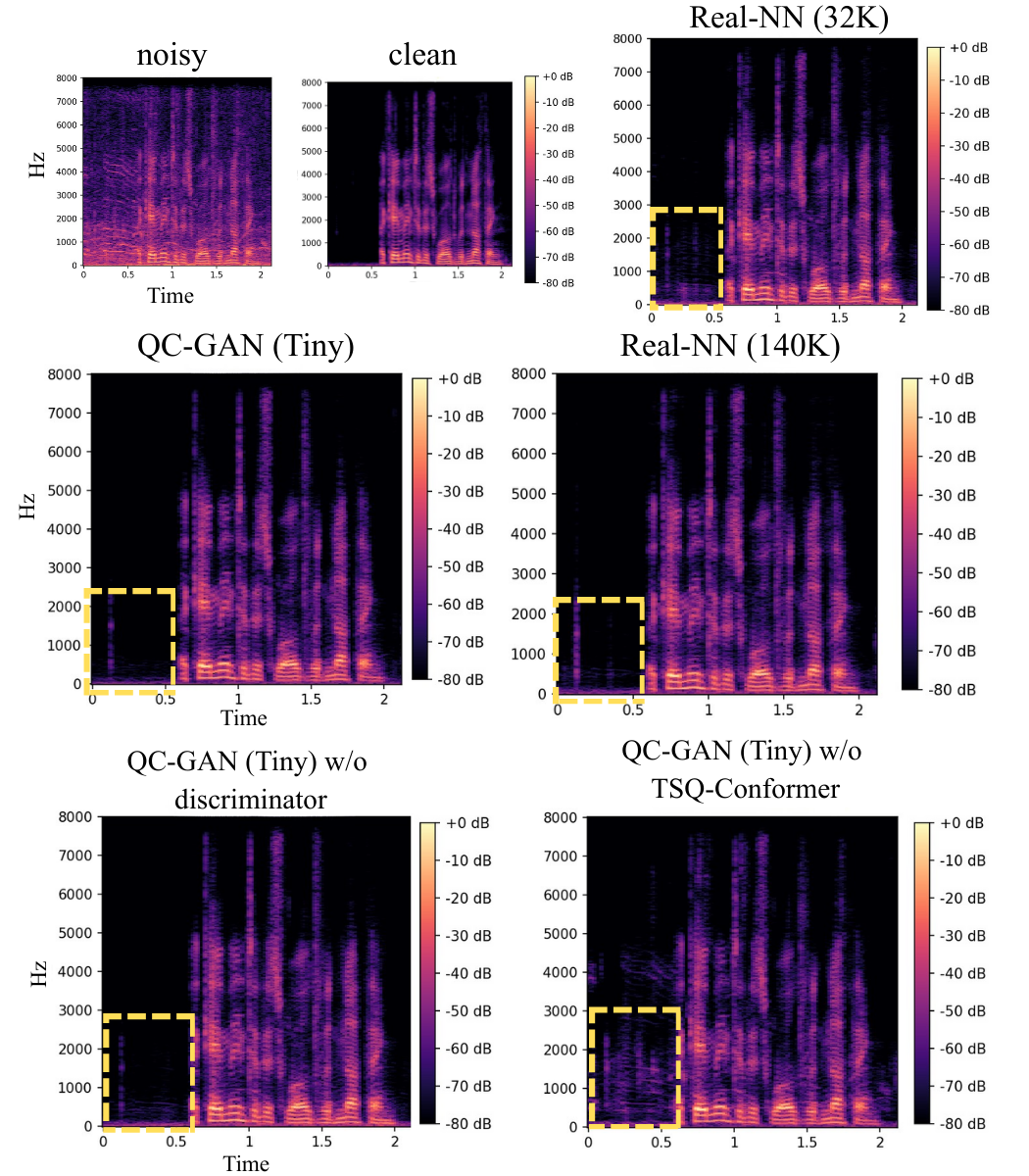}
    \caption{Spectrogram comparison on a VoiceBank+DEMAND test sample. Yellow dashed boxes mark the pre-speech silent region (0--0.5\,s), where QC-GAN (Tiny) achieves noise suppression closest to the clean reference.}

    \label{fig:spectrogram}
\end{figure}
 
To validate the effectiveness of our architectural design choices, we conducted an ablation study using QC-GAN (Tiny) configured on the VoiceBank+DEMAND dataset~\cite{valentinibotinhao16_ssw}. To isolate the contribution of quaternion algebra, we constructed a real-valued counterpart, Real-NN, by replacing all quaternion layers with standard real-valued layers. We configured two variants: Real-NN (32K), which approximately matched the parameter count of QC-GAN (Tiny, 35K), and Real-NN (140K), which had $4\times$ the parameters, corresponding to the theoretical reduction ratio of the Hamilton product. For a fair comparison, both variants used the same four-channel input features (Eq.~\ref{eq:q-features}), training configuration, and loss functions as QC-GAN. Additionally, to assess the contribution of individual components, we evaluated the variants in which either the discriminator or the Conformer bottleneck was removed. These ablations were conducted for both QC-GAN and Real-NN (140K), enabling a direct comparison of how each component interacts with the quaternion versus real-valued representations. The results are summarized in Table~\ref{tab:ablation_quaternion}.

\subsection{Efficacy of Quaternion Representation}
As shown in Table~\ref{tab:ablation_quaternion}, QC-GAN (35K) outperformed the parameter-matched Real-NN (32K) across all metrics, with STOI on par (PESQ: 3.12 $\rightarrow$ 3.23, COVL: 3.73 $\rightarrow$ 3.79). 
Moreover, Real-NN (140K), which used $4\times$ the parameters corresponding to the theoretical reduction ratio of the Hamilton product, attained a PESQ of 3.29. QC-GAN achieved a competitive PESQ of 3.23 with only 25\% of the parameters, confirming that quaternion algebra effectively maintained the representational capacity of a real-valued network at a fraction of the parameter cost.
 
Figure~\ref{fig:pesq_curve} reveals further distinctions in the learning dynamics. QC-GAN converged to a stable plateau after 30--40 epochs and maintained a consistent performance thereafter. In contrast, Real-NN (32K) fluctuated at a lower level throughout training, suggesting that the limited capacity of the real-valued model was insufficient to learn stable cross-component relationships. Real-NN (140K) achieved both higher PESQ and comparable training stability to QC-GAN; however, this required $4\times$ the parameters. QC-GAN attains competitive performance with only 25\% of those parameters, demonstrating that quaternion algebra’s inductive bias, which structurally couples the magnitude and phase components during feature transformation, achieved a favorable trade-off between performance and parameter efficiency.
Figure~\ref{fig:spectrogram} shows that QC-GAN achieved noise suppression in the pre-speech silent region closest to the clean reference among the compared models.

\subsection{Effect of Discriminator}
Removing the discriminator degraded COVL in both architectures (QC-GAN: 3.79 $\rightarrow$ 3.76; Real-NN 140K: 3.90 $\rightarrow$ 3.76), confirming its overall contribution to perceptual quality. For QC-GAN, the discriminator yielded a clear PESQ gain (3.14 $\rightarrow$ 3.23) at the cost of a slight CBAK decrease (3.40 $\rightarrow$ 3.36), suggesting that adversarial training steered the generator toward perceptually salient improvements beyond those achieved by reconstruction losses alone. 
In Real-NN (140K), removing the discriminator slightly improved PESQ (3.29 $\rightarrow$ 3.32) while degrading COVL (3.90 $\rightarrow$ 3.76), suggesting that the discriminator primarily contributed to the overall perceptual quality (COVL) rather than PESQ optimization in real-valued networks. In contrast, QC-GAN benefited from the discriminator in both PESQ and COVL, indicating that adversarial training is more consistently effective under quaternion representations.
 
\subsection{Effect of Conformer}
Removing the Conformer bottleneck causes substantial degradation in both architectures: QC-GAN dropped from 3.23 to 2.99 ($-$0.24) and Real-NN (140K) from 3.29 to 3.10 ($-$0.19). Although this removal also reduces the parameter count, the disproportionately large degradation compared with other ablations (e.g., discriminator removal) indicates that the attenuated long-range modeling capability is the dominant factor.
Notably, QC-GAN exhibited greater degradation than the Real-NN, despite having fewer parameters. This could be attributed to the fact that quaternion convolutions, while effective at capturing local inter-component relationships, rely more heavily on the Conformer's self-attention to model long-range temporal dependencies. Importantly,  both architectures exhibit substantial degradation in both cases, confirming that the Conformer's long-range modeling capability is preserved when reformulated entirely within quaternion algebra. This validates the Q-Conformer as a viable quaternion counterpart to the standard Conformer, as it retains the essential role of self-attention for speech enhancement under the structural constraints of Hamilton’s product.

\section{Discussion}
This section examines the advantages of QC-GAN over Real-NN, focusing on the learning characteristics and phase handling that contribute to its superior parameter efficiency and improved performance. The comparison uses QC-GAN (Tiny) and Real-NN from the ablation study.
 
\subsection{Why Quaternion Networks Achieve Parameter Efficiency}
Quaternion layers implement inter-component mixing via the Hamilton product, where four real-valued sub-matrices are reused to generate all quaternion components (Eq.~\ref{eq:qfc_matrix}). This yields a substantial parameter reduction compared with an equivalent real-valued layer operating on four channels. For a mapping from $C_{\text{in}}$ to $C_{\text{out}}$ quaternion channels (kernel size $K$), a real-valued layer requires $16C_{\text{out}}C_{\text{in}}K$ parameters, whereas a quaternion layer requires only $4C_{\text{out}}C_{\text{in}}K$ parameters, i.e., a $4\times$ reduction. 
Beyond the parameter count, the Hamilton product imposes a structured $4\times4$ block mixing pattern (Eq.~\ref{eq:qfc_matrix} and Figure~\ref{fig:NN}) rather than an unconstrained inter-channel transformation, enabling rotation-like coupling that processes the four components: spectral dynamics $\Delta\log(|Y|_\epsilon)$, static magnitude $\log|Y|$, and phase $(\cos\theta_Y,\sin\theta_Y)$
(Eq.~\ref{eq:q-features}) as a unified representation. This structure aligns well with speech signals, where magnitude and phase are inherently coupled, and the $(\cos\theta_Y,\sin\theta_Y)$ pair preserves phase circularity.
 
This inductive bias is particularly beneficial in the ultra-compact regime (35K parameters). A 4-channel Real-NN must allocate limited capacity to learn consistent cross-channel relationships from data under a highly unconstrained mixing space, which can be parameter-inefficient. In contrast, QNNs encode cross-component coupling directly through the Hamilton product, allowing the model to focus on time--frequency patterns relevant to enhancement.

\subsection{Phase Preservation Analysis}
 
\begin{figure}[t]
\centering
\includegraphics[width=0.85\linewidth]{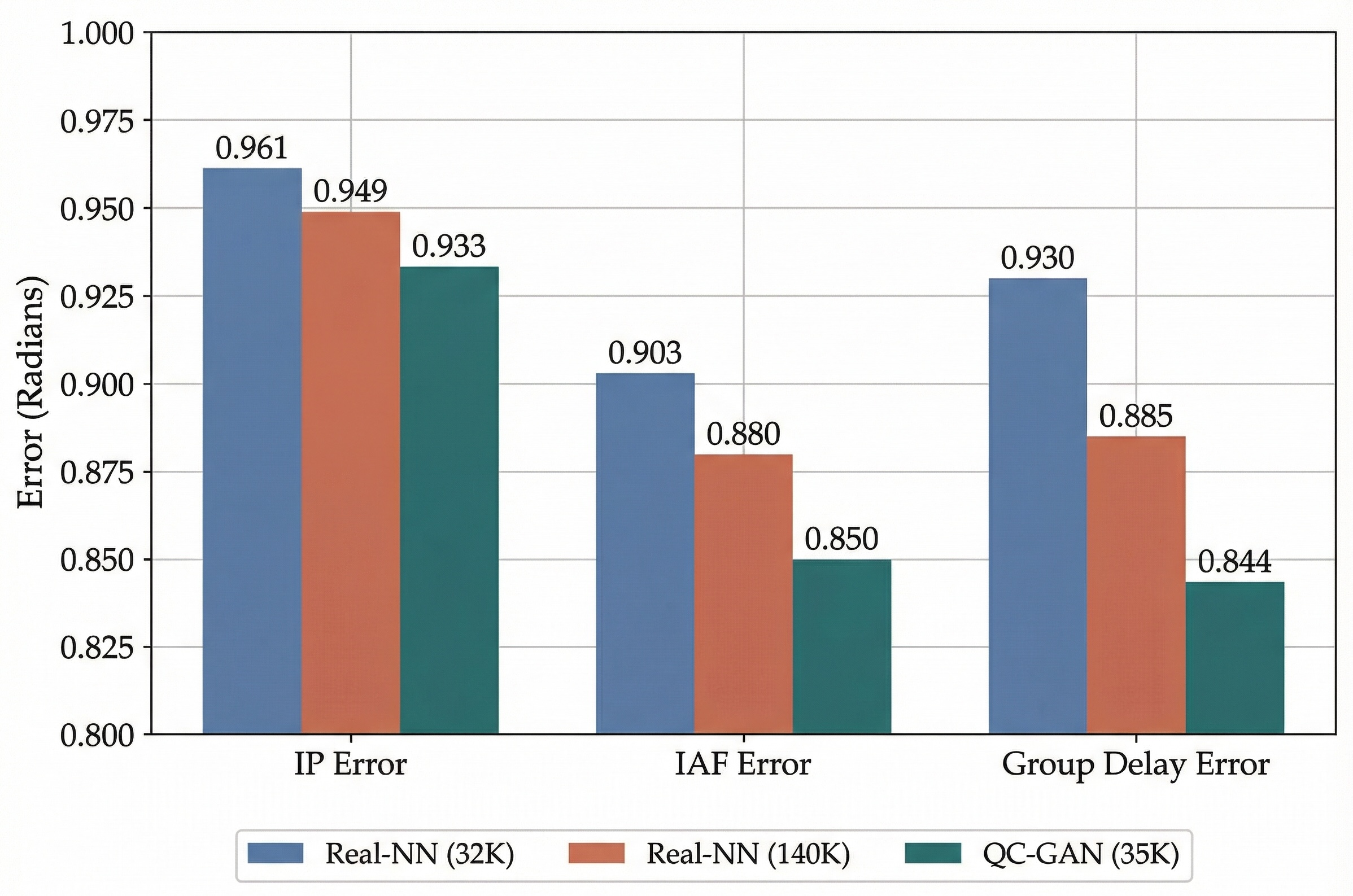}
\caption{Phase reconstruction metrics on the VoiceBank+DEMAND test set. Lower values indicate better phase preservation. QC-GAN outperforms both Real-NN variants across all three metrics.}
 
\label{fig:phase_metrics}
\end{figure}
 
\begin{figure}[t]
\centering
\includegraphics[width=\linewidth]{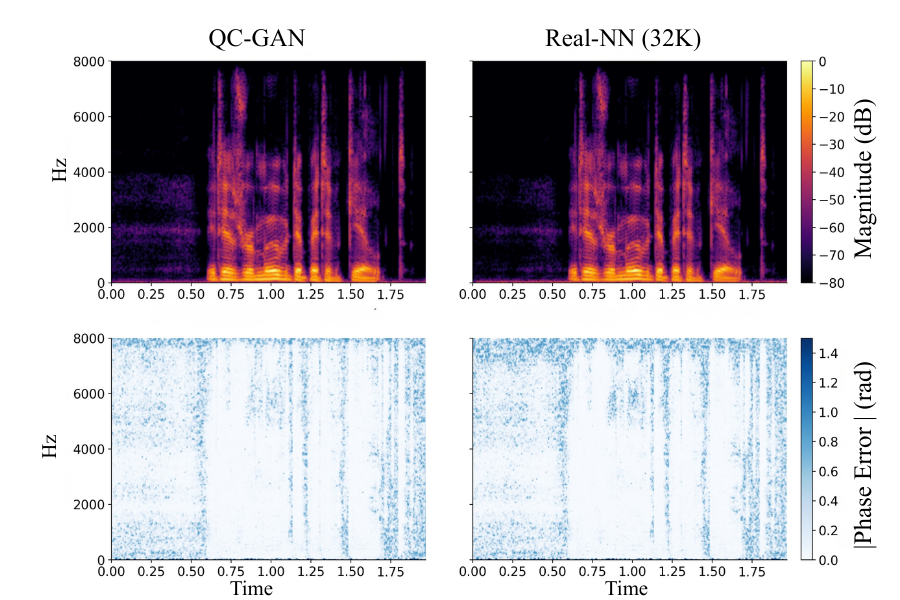}
\caption{Phase reconstruction for QC-GAN and Real-NN (32K): spectrograms (top) and absolute phase error weighted by clean speech amplitude (bottom). QC-GAN reduces mean weighted phase error by 14.7\% (0.414$\to$0.353\,rad).}
 
\label{fig:Phase_Error}
\end{figure}
 
The ablation study (Table~\ref{tab:ablation_quaternion}) shows that QC-GAN outperformed the parameter-matched Real-NN (32K) in PESQ and achieved competitive performance with Real-NN (140K), which uses $4\times$ the parameters, despite the identical training setup. To investigate whether this advantage originated from improved phase reconstruction, we evaluated the phase-related metrics on the entire VoiceBank+DEMAND test set (824 utterances). Specifically, we measured the Instantaneous Phase (IP), Instantaneous Angular Frequency (IAF), and Group Delay (GD) errors~\cite{ai2023neural}, all of which capture perceptually relevant aspects of phase accuracy.
 
As shown in Figure~\ref{fig:phase_metrics}, QC-GAN consistently reduced phase errors compared with the Real-NNs (32K, 140K), achieving relative reductions of 2.91\% in IP (0.961$\rightarrow$0.933 rad), 5.87\% in IAF (0.903$\rightarrow$0.850 rad), and 9.25\% in GD (0.930$\rightarrow$0.844 rad). Notably, the improvement was most pronounced for GD, indicating more accurate phase slopes across frequencies and improved temporal consistency of reconstructed speech. 
This improvement is also seen in the bottom row of Figure~\ref{fig:Phase_Error}, where the weighted absolute phase error map of QC-GAN exhibits noticeably lighter colors than that of Real-NN (32K), demonstrating superior phase reconstruction across the time-frequency plane. Furthermore, Figure~\ref{fig:spectrogram} shows that QC-GAN achieves cleaner noise suppression in the pre-speech silent region than Real-NN (140K), despite the latter achieving a slightly higher PESQ, suggesting that improved phase consistency contributes to local noise suppression.
Overall, these results support our hypothesis that Hamilton-product-based coupling helps maintain consistent relationships between magnitude and the $(\cos\theta,\,\sin\theta)$ phase representation, leading to more stable phase reconstruction under an ultra-compact parameter budget.

\subsection{Computational Trade-off}
 
\begin{table}[t]
\caption{RTF comparison. MACs are real-valued, for 1\,s of audio.
CPU: Intel Xeon Gold 6342 (4 threads); GPU: NVIDIA A100 80GB PCIe.
For each quaternion model, the value in parentheses represents the qMAC count of
the parametrized layers ($1$ qMAC $=16$ real MACs); $^{\dagger}$the
real-MAC value of QC-GAN (Tiny) additionally includes attention.}
  \label{tab:macs_rtf}
  \centering
  \small
  \resizebox{\columnwidth}{!}{%
  \begin{tabular}{l|ccc}
    \toprule
    \textbf{Model}  & \textbf{MACs(G)} & \textbf{CPU RTF} & \textbf{GPU RTF} \\
    \midrule
    Real-NN (32K)    & 0.07 & 0.106 & 0.0033 \\
    Real-NN (140K)   & 0.42 & 0.179 & 0.0092 \\
    \midrule
    QC-GAN (Tiny)    & 3.75 (0.22)$^{\dagger}$ & 0.890 & 0.015 \\
    \quad w/o TSQ-Conformer  & 2.96 (0.19) & 0.107 & 0.012 \\
    \bottomrule
  \end{tabular}
  }
\end{table}
 
Although quaternion operations deliver superior perceptual quality, they incur computational overhead. As shown in Table~\ref{tab:macs_rtf}, QC-GAN (Tiny) requires 3.75\,G real-valued MACs (0.22\,G qMACs from parametrized layers, plus attention, where $1$ qMAC $=16$ real MACs) compared with 0.07\,G for Real-NN (32K). However, QC-GAN (Tiny) achieves a CPU RTF of 0.89 on a 4-thread CPU, remaining below the real-time threshold (RTF $<1.0$). Profiling showed the TSQ-Conformer, rather than the total MAC count, dominated CPU runtime: removing this module reduced the CPU RTF to 0.107. Moreover, the near-identical CPU RTFs of QC-GAN (Tiny) w/o TSQ-Conformer and Real-NN (32K) (0.107 vs.\ 0.106), despite the large MAC gap, indicate that the runtime is constrained by dispatch overhead rather than arithmetic complexity. All variants achieved GPU RTFs well below 1.0 (0.015 for QC-GAN Tiny), confirming real-time deployability. Nonetheless, the attention bottleneck leaves limited CPU margin, which can be improved through fused quaternion kernels or linear attention.

\section{Conclusion}
We proposed QC-GAN, a speech-enhancement framework that integrates Quaternion Neural Networks into a Conformer architecture with MetricGAN-based training. Leveraging the Hamilton product, QC-GAN achieved performance competitive with SoTA models with approximately half their parameters, and the Tiny model (35K) attained a PESQ of 3.23 on VoiceBank+DEMAND, outperforming existing lightweight methods. Evaluation on the DNS Challenge confirmed generalization to real-world conditions.
Ablation studies showed that quaternion representations improved phase reconstruction---reducing group delay error by approximately 9\%---confirming phase preservation as the key mechanism underlying the framework's performance advantage.

Future work includes reducing inference latency through fused quaternion operators and linear attention, extending the framework to multi-channel and streaming settings, and exploring alternative assignments of spectral and phase information to the quaternion components.

\newpage
 
\section{Acknowledgements}
The authors thank the anonymous reviewers for their constructive comments.

\section{Generative AI Use Disclosure}
Anthropic Claude was used for English translation assistance, and Paperpal (by Editage) was used for grammatical proofreading. The authors reviewed and edited the outputs and take full responsibility for the final content of this paper.
 
\bibliographystyle{IEEEtran}
\bibliography{main}

\end{document}